# Density-controlled growth of vertical InP nanowires on Si(111) substrates

A. Jaffal[1,2], P. Regreny[2], N. Chauvin[1] and M. Gendry[2]

[1]Université de Lyon, Institut des Nanotechnologies des Lyon-INL, UMR 5270 CNRS, INSA de Lyon, 7 avenue Jean Capelle, 69621 Villeurbanne cedex, France
[2]Université de Lyon, Institut des Nanotechnologies des Lyon-INL, UMR 5270 CNRS, Ecole Centrale de Lyon, 36 avenue Guy de Collongue, 69134 Ecully cedex, France

E-mail: ali.jaffal@insa-lyon.fr

**Abstract**

A procedure to achieve the density-controlled growth of gold-catalyzed InP nanowires (NWs) on (111) silicon substrates using the vapor-liquid-solid method by molecular beam epitaxy is reported. We develop an effective and mask-free method based on controlling the number and the size of the Au-In catalyst droplets in addition to the conditions for the NW nucleation. We show that the NW density can be tuned with values in the range of 18 $\mu m^{-2}$ to < 0.1 $\mu m^{-2}$ by the suitable choice of the In/Au catalyst beam equivalent pressure (BEP) ratio, by the phosphorous BEP and the growth temperature. The same degree of control is transferred to InAs/InP quantum dot-nanowires, taking advantage of the ultra-low density to study by micro-photoluminescence the optical properties of a single quantum dot-nanowires emitting in the telecom band monolithically grown on silicon. Optical spectroscopy at cryogenic temperature successfully confirmed the relevance of our method to excite single InAs quantum dots on the as-grown sample, which opens the path for large-scale applications based on single quantum dot-nanowire devices integrated on silicon.

## 1. Introduction

Semiconductor Nanowires (NWs) have witnessed an explosion of interest in the fields of optoelectronic devices [1] and solar cells [2]. In particular, InP and GaAs NWs are the building block materials to explore such fields because of their remarkable optical and electronic properties as well as the advanced research concerning their growth. Molecular beam epitaxy (MBE) is one of the most suitable growth technique to grow Au-catalyzed or self-catalyzed III-V NWs using the Vapor-Liquid-Solid (VLS) method [3][4]. This method ensures the epitaxial relationship with the substrate and therefore paves the way for the monolithic integration of III-V NWs on Silicon (Si) substrates thanks to their small footprint and diameter which allow the NWs to accommodate the mismatch-induced strain with elastic relaxation by their free facets [5][6]. In such a scenario, the development of III-V NWs-on-Si has attracted a great interest towards realizing complementary-metal-oxide-semiconductor (CMOS) compatible devices for various applications, especially light-emitting devices such as single photon sources (SPSs) [7] or nanolasers [8]. One major challenge to reach monolithic NW-based SPSs on Si is to grow low NW density on Si allowing to optically excite a single NW. In this context, tuning the density of the NWs grown in a self-organized manner on silicon substrates is still a remaining challenge and achieving it would offer a simple, low cost and mask-free pathway for large-scale applications.

One way to tune the NW density is by substrate patterning followed by selective area growth (SAG) [9]. However, this method demands considerable efforts in terms of time, expertise and equipment. In an alternative approach to control the NW density on Si is the use of droplet epitaxy technique [10]. Although this method is efficient to control the density and diameter of Ga droplets that initiate the GaAs NWs, the transition yield from droplets to NWs is not 100%. In a more recent work, it was shown that the density of Ga catalyzed NWs can be controlled by a careful choice of Ga flux and growth temperature [11]. As we are interested in achieving NW based SPSs emitting in the telecom band, we know that InAs QDs, as single photon emitters, are usually embedded either in InP NWs [12] or in GaAs NWs [13]. The latter seems to be limited due to the strain induced by the GaAs NW on the InAs QD [14] while growing self-catalyzed InP NWs by VLS-MBE is not an easy task and in the successful cases the crystalline structure of the InP NWs is faulty full with a zinc blende-wurtzite (ZB-WZ) phase mixture [15]. In this case, the QD-NW heterostructures would feature structural defects and low-quality interfaces in the QD vicinity. Therefore, the chances to reach SPSs with such NWs are reduced where it is necessary to grow the QD in defect-free NWs [16][17].

In the case of Au-catalyzed InP NWs grown by VLS-MBE, pure WZ NWs can be grown by controlling the growth temperature and the V/III flux ratio [18–20]. Consequently, the Au-catalyzed InP NWs are the main structures used as waveguides to guide the QD emitted light for efficient single photon source at telecom wavelengths [21–24]. The density control of Au catalyzed NWs has been reported in the work of Wang et al. [25]: the ZnO NW density was modified by varying the Au thin film thickness with a drop in the NW density from ~ 110 to ~15 μm$^{-2}$ when the Au film thickness was increased from 1 to 8 nm. Moreover, Xu et al.[26] have studied the effect of annealing temperature on the Au film, in parallel to the Au film thickness, to grow GaAs NWs on GaAs substrates by metal organic chemical vapor deposition (MOCVD). In both studies, it is shown that the size of the Au droplets increases as the Au film thickness increases with a drop in the Au droplets density and consequently a strong decrease of the NW density. Despite the fact that this method is effective when the Au thin film is evaporated on GaAs substrates, such an approach is limited when Au is deposited on Si(111) substrates. This is due to the formation of Au-Si alloy which, due to the presence of an eutectic point at 360°C in the Au-Si diagram,[27] will be liquid at this temperature. It is therefore associated with a digging of the substrate below the formed droplets and three other (111) planes, inclined at 19° with respect to the horizontal (111) surface of the substrate can then be revealed [28]. This issue can be avoided by the use of Au-In droplets as catalyst to limit the interaction of Au with Si and thus allow the growth of vertical InP NWs on Si(111). In the recent of work of Mavel et al. [18], it has been shown that the Au-In droplets deposited on Si(111) at 500°C will form an Au-In core-shell structure that will result in a vertical yield of InP NWs ~ 100%. However, no control over the InP NW density has been reported.

In this paper, we experimentally demonstrate the control over the InP NW density using Au-In droplets as catalysts in-situ deposited at 500°C on a Si(111) substrate using solid source molecular beam epitaxy (ss-MBE). Firstly, we show the impact of the In/Au flux ratio used for the formation of the catalyst droplets on the NW density. Then, we studied the influences of the phosphorous ($P_2$) pressure and the growth temperature $T_G$ on the NW density. We show that the NW density can be thus tuned from ~ 18 μm$^{-2}$ to < 0.1 μm$^{-2}$. This allowed us to perform single InAs/InP QD-NW photoluminescence spectroscopy on the as-grown sample opening a novel method for the monolithic growth of III-V QD-NWs SPS devices on Si substrates.

## 2. Experimental methods

The InP NW samples were grown on Si(111) substrates using the VLS assisted solid source-MBE with Au-In droplets as catalyst. We have used the same substrate preparation and Au-In droplet formation procedures mentioned in a previous work [18]. In this work, we have shown that the deposition temperature of 500°C was the optimal temperature to reach high vertical yield of InP NWs [18]. Therefore, the Au-In droplets were in-situ formed at 500°C by co-evaporating Au and In during 60 sec. After the droplet formation, the sample temperature was decreased to the NW growth temperature $T_G$ in the 360°C-380°C range. The InP NWs were grown during 14 min with an In beam equivalent pressure (BEP) equal to $6 \times 10^{-7}$ torr corresponding to a growth rate of 1 ML/s for an InP layer and a V/III BEP ratio in the 10 – 20 range.

The grown samples were observed by scanning electron microscopy (SEM) to have access to the density ($d_{NW}$) of the InP NWs in addition to their morphological properties. The NW density was measured at three different positions for each sample of ~ 1 cm$^2$ surface area to verify that the NW density is homogenous all over the sample.

The samples were optically characterized by photoluminescence (PL) and micro-photoluminescence (μPL). The samples were mounted inside a close cycled helium cooled cryostat allowing measurements at T = 12 K.

The light emitted from the NWs was collected and analyzed by a liquid nitrogen cooled Indium Gallium Arsenic (IGA) detector camera coupled to a monochromator. Continuous wave (cw) PL spectroscopy was carried out using a 532 nm diode-pumped solid-state laser source with a ≈ 200 µm laser spot size. µPL was performed using a cw single mode fiber-coupled 780 nm laser source. The light was focused on the sample using a x50 microscope objective with a numerical aperture NA = 0.4. The laser spot on the sample was around 3-4 µm in diameter allowing optical spectroscopy measurements on single NWs.

## 3. Results and discussion

### 3.1 Effect of the In/Au BEP ratio on the NW density

Our first approach to control the density of the InP NWs was based on modifying the In/Au BEP ratio for the catalyst droplet formation. The deposition of Au-In droplets was performed using an In BEP equal to $2.5 \times 10^{-7}$ torr, corresponding to a growth rate of 0.2 ML/s for an InP layer whereas the Au BEP was varied between $1.8 \times 10^{-8}$ torr and $4.2 \times 10^{-9}$ torr. Thus, a series of 5 InP NW samples were grown with an In/Au BEP ratio = 14, 17, 20, 25, and 58, respectively. As already shown in [18], during this step Au (core) - In (shell) droplets are formed on the Si substrate (Figure 1a). Then, prior to the growth of the InP NWs, when the $P_2$ valve is opened during 10 sec InP pedestals are formed with an Au droplet on top of them as illustrated in Figure 1b. According to [18], the formation of the InP pedestals is due to the transformation of the In liquid shell of the Au-In droplets into solid InP once exposed to the $P_2$ flux. During the formation of the InP pedestal, the Au core will migrate from the core to the top of the pedestal and thus it will promote vertical InP NWs due to the presence of the Au droplets on top of an InP(111) surface.

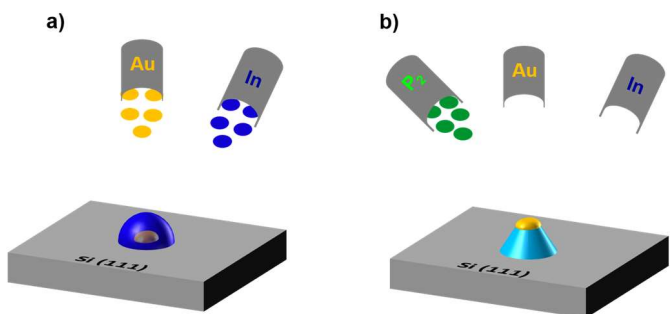

**Figure 1.** Schematic representation of : a) an Au-In core-shell droplet deposited at 500°C on Si(111) and b) an InP pedestal with an Au droplet on top after opening the $P_2$ valve during 10 sec (from ref [18]).

The InP NWs were then grown at $T_G$ = 380°C with a V/III BEP ratio = 16. Figure 2a-f shows the tilted view (45°) SEM images of the as-grown NWs with In/Au BEP ratio = 14, 17, 20, 25, and 58, respectively. The variation of the NW density as a function of the In/Au BEP ratio is shown in Figure 2g. Clearly, InP NWs catalyzed using In/Au BEP ratio = 14 leads to the highest NW density ~ 18 µm$^{-2}$. Then, as the In/Au BEP ratio increases, i.e. the Au BEP is lower, the NW density drops gradually to reach a value as small as ~ 0.1 µm$^{-2}$. Despite the large difference in the NW density as a function of the In/Au BEP ratio, all InP NWs exhibit very similar diameter, $D_{NW}$, in the 50-65 nm range and length, $L_{NW}$, in the 1-1.2 µm range (Figure 2g).

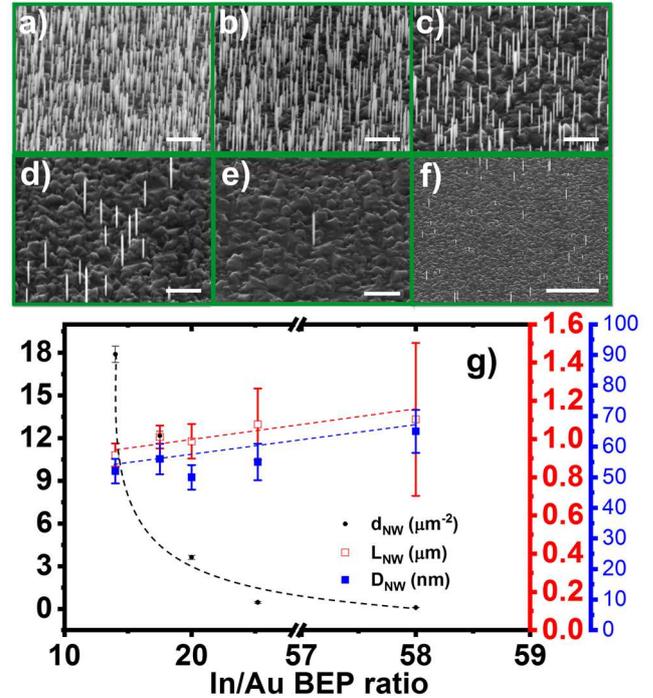

**Figure 2.** Tilted (45°) SEM images showing the variation of the InP NW density as a function of the In/Au BEP ratio used for the droplet formation: In/Au BEP ratio = a) 14, b) 17, c) 20, d) 25, e) 58. f) SEM image of the same sample as in (e) at low magnification confirming the ultra-low NW density. g) The evolution of the InP NW density (black), length (red) and diameter (blue) as a function of the In/Au BEP ratio. The dashed lines in (g) are guidelines for the eyes. Scale bars in (a-e) 1 µm and (f) 5 µm.

To understand the obtained results shown in Figure 2, we have studied the droplets formed with an In/Au BEP ratio = 14 and 58 (Figure 3). For the In/Au BEP ratio = 14, corresponding to an Au BEP = $1.76 \times 10^{-8}$ torr, a high density of small Au-In droplets distributed in two populations is observed (Figure 3a). The first population consists of droplets with diameter < 20 nm and a high density (~600 µm$^{-2}$) and the second population consists of droplets with diameter in the 20-40 nm range and a lower density (~ 300 µm$^{-2}$). On the other hand, when the In/Au BEP ratio = 58, for Au BEP = $4.2 \times 10^{-9}$ torr, the formed droplets exhibit three populations: two populations of droplets are similar to what is obtained with In/Au BEP ratio = 14, with diameter < 20 nm and in 20-40 nm range, and with densities of ~$2.7 \times 10^{3}$

μm$^{-2}$ and ~1-2 μm$^{-2}$ range, respectively (Figure 3b). The third population consists of droplets with diameter in the 400-500 nm range and a very low density (~0.1 μm$^{-2}$) (Figure 3c).

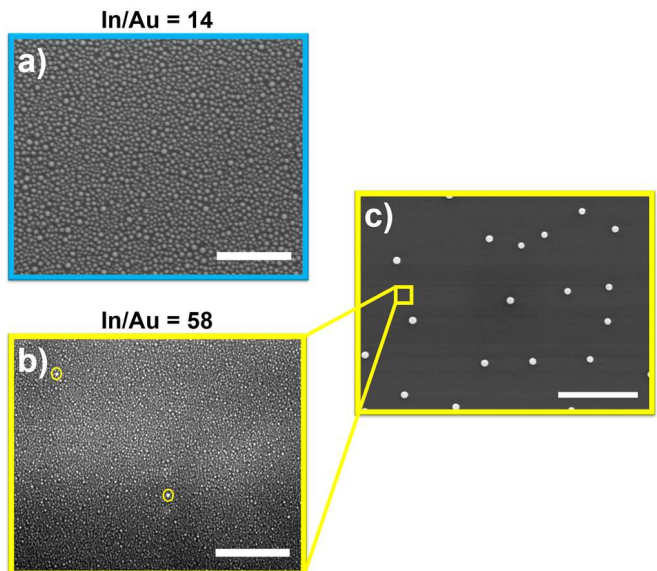

**Figure 3.** Top view SEM images of the Au-In droplets deposited at 500°C with an In/Au BEP ratio equal to: a) 14, b) 58 with the droplets in the yellow circles in the 20-40 nm diameter range. c) Low magnification image of the sample in (b) showing the droplet formation with diameter in the 400-500 nm range. Scale bars in (a), (b) 500 nm and (c) 5 μm.

The differences in the size distributions and densities of the Au-In droplets can be explained taking into account the diffusion length of the In adatoms on Si substrate during the co-evaporation of Au and In. When a high amount of Au is used to form the droplets (In/Au = 14), the Au atoms will limit the diffusion of the In adatoms and thus it will result in a high density of Au-In droplets with small diameters. Contrariwise, for a lower amount of Au atoms (In/Au = 58), the diffusion length of the In adatoms will be longer and as a consequence the size distribution of the Au-In droplets will be modified with moreover the formation of a third population of larger droplets. These dramatic modifications in the Au-In droplet density and size distribution, as summarized in the histogram shown in Figure 4a and b, are the reason behind the strong variation in the density of the grown InP NWs. In Figure 4c and d, we represent schematically the different populations of the Au-In droplets for In/Au BEP ratio = 14 and 58, respectively. We will divide them into A, B and C populations with the A population corresponding to Au-In droplets < 20 nm in diameter, the B population corresponding to droplets with diameter in the 20-40 nm range and the C population corresponding to droplets with diameter in the 400-500 nm range.

To know which droplet population contributes to the growth of the InP NWs, we have performed a statistical study on the density of the different droplet populations ($d_{droplets}$) and compared it to the NW density ($d_{NWs}$) at two different In/Au BEP ratios = 14 and 58 (Table 1). For Au-In droplets with diameter < 20 nm (A population), the density of the droplets is high (600 to 2.7x10$^3$ μm$^{-2}$) for both In/Au BEP ratios however a large difference between the density of those droplets and that of the InP NWs is observed. This is a first evidence that these droplets haven't promoted the growth of the InP NWs. Concerning the droplets with diameter in 20-40 nm (B population), the $d_{droplets}/d_{NWs}$ values for In/Au BEP ratio = 14 and 58 are lower with values of ~15 and in the 10-20 range, respectively. This can be an evidence that the droplets with diameter in the 20-40 nm range are responsible for the growth of InP NWs. These results are in good agreement with the work in [18] where it was shown that the Au-In droplets 35-45 nm in diameter leads to the growth of the InP NWs. The remaining C population is the big Au-In droplets with diameter in the 400-500 nm range that are formed only in the case of In/Au BEP ratio = 58. The density of these droplets is very similar to that of InP NWs and we would easily attribute the growth of low density InP NWs to such droplets but this is not the case. To assert that, we have performed the growth of InP NWs with In/Au BEP ratio = 58 for a short time (2 min) in order to see the early stages of the NW growth with such big droplets. The SEM image in Figure 5 shows that these droplets form 3D islands of InP.

In summary on this part, we believe that the Au-In droplets resulting in the nucleation and growth of the InP NWs are the ones with diameter in the 20-40 nm range (B population). By modifying the In/Au BEP ratio we can therefore modify the density of these droplets, and thereafter achieve an important tuning of the NW density.

### 3.2 Effect of the phosphorous BEP and the growth temperature on the NW density

After the first results showing the strong impact of the In/Au BEP ratio on the as-grown InP NW density, we have further investigated the influence of the phosphorous BEP ($P_{P2}$), or of the V/III BEP ratio, and the growth temperature ($T_G$). For this study, we have chosen the two extreme cases with In/Au BEP ratio = 14 and 58. Figure 6 shows the SEM images of the as-grown InP NWs. The characteristics of the InP NWs are summarized in Table 2.

First, the growth temperature of InP NWs was fixed at 360°C and the $P_2$ BEP was varied to 6x10$^{-6}$ torr and 1.2x10$^{-5}$ torr, corresponding to a V/III BEP ratio = 10 and 20, respectively. Whatever the In/Au BEP ratio, we can observe a slight decrease in $d_{NW}$ when the $P_2$ BEP is increased: from 18 μm$^{-2}$ to 13 μm$^{-2}$ for In/Au BEP ratio = 14 (Figure 6a and

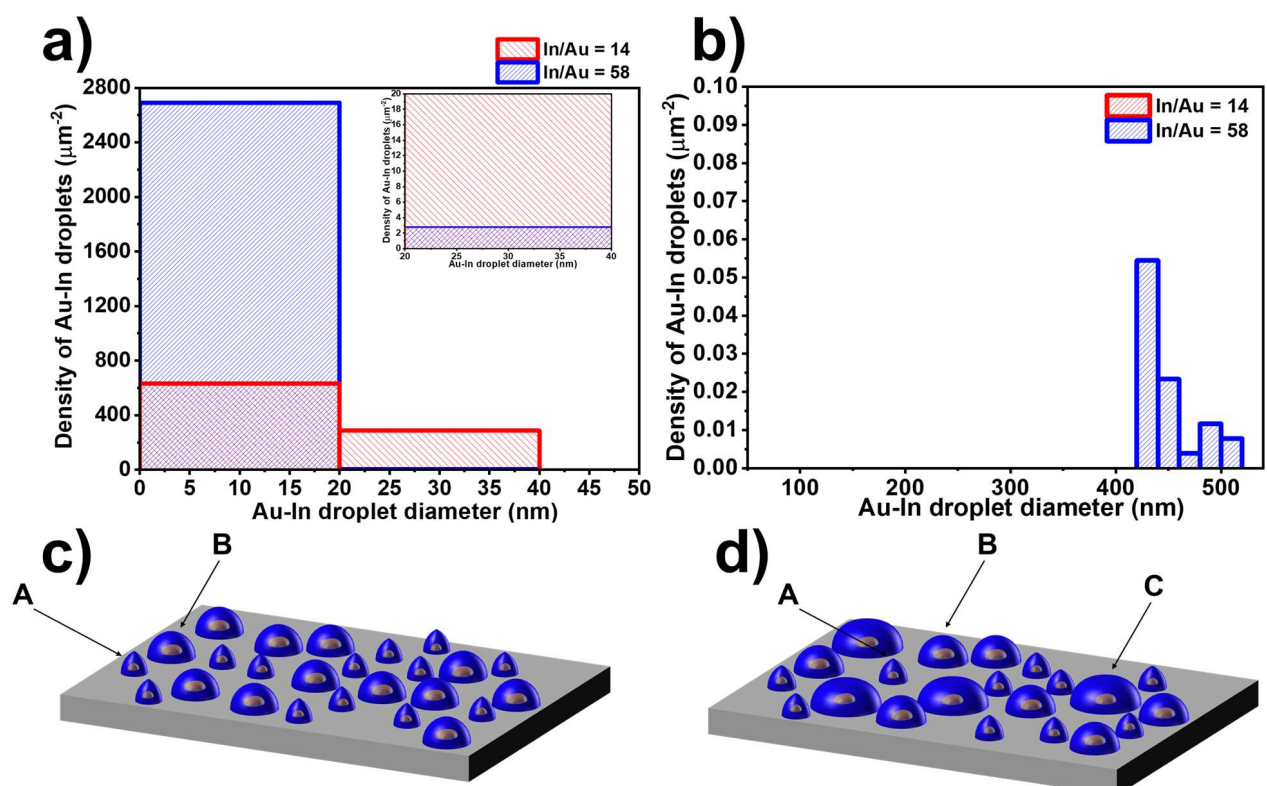

**Figure 4.** a) Histogram showing the diameter distribution in the 0-50 nm range of the Au-In droplets with In/Au BEP ratio = 14 and 58, respectively, as measured on the SEM images of Figure 3(a) and (b). Inset is a magnification for the Au-In droplets with diameter in the 20-40 nm range. b) Histogram showing the diameter distribution in the 400-500 nm range of the Au-In droplets for In/Au BEP ratio = 58, as measured on the SEM image of Figure 3(c). (c, d) Schematic representations of the Au-In droplets with two and three different populations for In/Au BEP ratio = 14 and 58, respectively.

| In/Au ratio | $d_{droplets\ (<20\ nm)}$ ($\mu m^{-2}$) | $d_{droplets(20-40\ nm)}$ ($\mu m^{-2}$) | $d_{droplets\ (400-500\ nm)}$ ($\mu m^{-2}$) | $d_{NW}$ ($\mu m^{-2}$) | $\dfrac{d_{droplets(<20\ nm)}}{d_{NW}}$ | $\dfrac{d_{droplets(20-40\ nm)}}{d_{NW}}$ |
|---|---|---|---|---|---|---|
| 14 | ~600 | ~300 | No droplets | ~20 | ~30 | ~15 |
| 58 | ~2.7x10$^3$ | 1-2 | ~0.1 | ~0.1 | ~2.7x10$^4$ | 10-20 |

**Table 1.** Comparison of the Au-In droplet density ($d_{droplets}$) of different diameters with the NW density ($d_{NWs}$) for In/Au BEP ratio = 14 and 58.

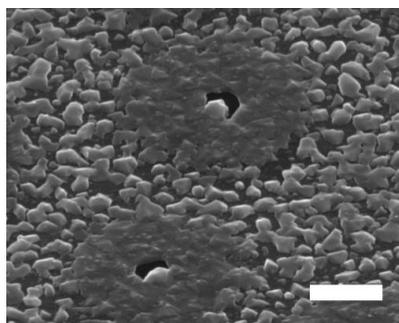

**Figure 5.** Tilted (45°) SEM image after 2 min growth of InP on a surface with droplets formed using the In/Au BEP ratio = 58, showing the formation of 3D islands of InP with the Au-In droplets having a diameter in the 400-500 nm range. Scale bar is 1 μm.

b) and from 3 μm$^{-2}$ to 2 μm$^{-2}$ for In/Au BEP ratio = 58 (Figure 6d and e).

Considering the mechanism shown in Figure 1, such a variation in the NW density can be explained by the behavior of the In shell-Au core droplets when they are subjected to the P$_2$ pressure for 10 seconds to lead to the formation of InP pedestals with an Au droplet being on their top [18]. The migration of the Au core is related to the P$_2$ pressure which induces the formation of InP from the In droplet shell, thus pushing the Au core to the top of InP pedestals. We can suppose that the higher is the P$_2$ pressure, the faster is the formation of InP pedestals, which could inhibit the Au core migration to the top of the pedestals and thus the nucleation and growth of the InP NWs. Therefore, the number of the InP pedestals with an Au droplet on their top could be tuned by the $P_2$ pressure and could have a non-negligible influence on the NW density. A noticeable change can be also observed in the NW length and diameter when modifying the $P_2$ pressure (see Table 2). The increase in the NW length and the decrease in the NW diameter for lower $P_2$ BEP (lower V/III BEP ratio), is explained by a higher In adatom diffusion on the NW facets towards the Au droplet when the $P_2$ pressure is lowered promoting a faster axial growth rate, In being a limiting source for the axial growth of Au-catalyzed In-based NWs [29]. Note however that we have previously shown in [19] that a low $P_2$ BEP corresponding to a V/III BEP ratio = 10 could favor the presence of stacking faults and thin ZB insertions in WZ InP NWs and therefore a V/III BEP ratio = 20 must be chosen to grow pure WZ NWs.

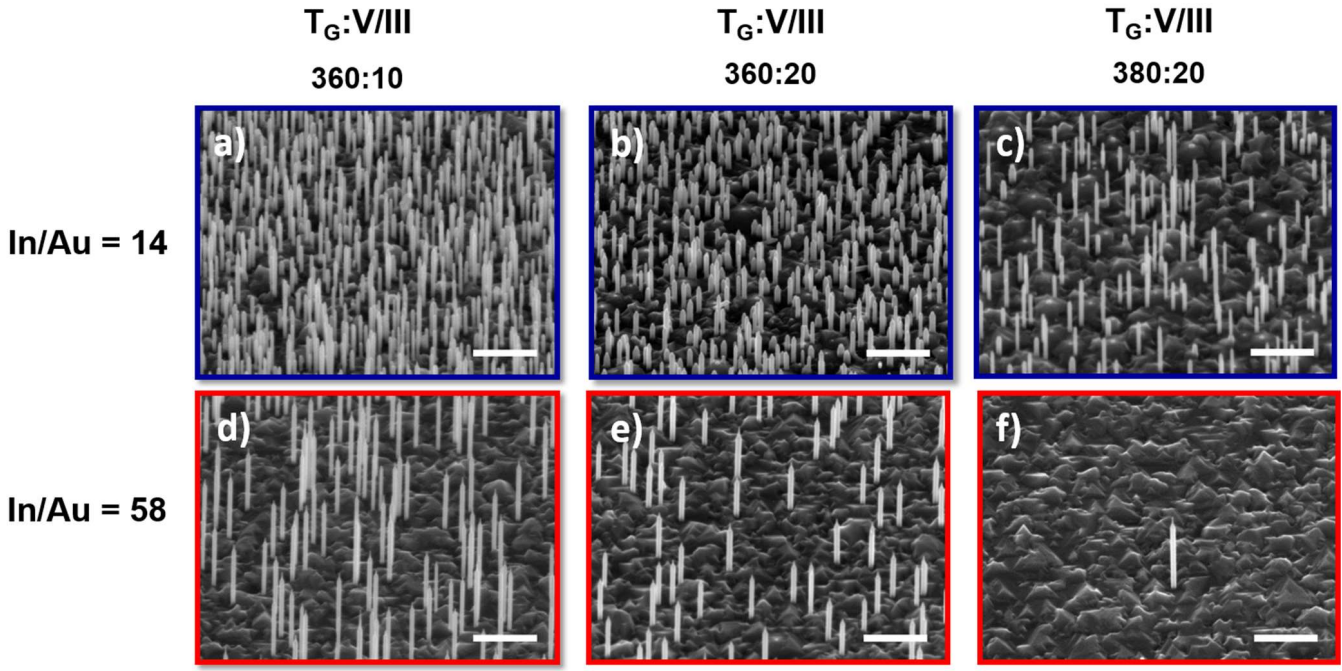

**Figure 6.** Tilted (45°) SEM images of the InP NWs grown at different growth temperature and $P_2$ BEP (or V/III BEP ratio), with In/Au BEP ratio = 14 (a to c) and In/Au BEP ratio = 58 (d to f). Scale bars are 1 μm.

|   | In/Au | $T_G$ (°C) | V/III | $d_{NW}$ (μm$^{-2}$) | $L_{NW}$ (μm) | $D_{NW}$ (nm) |
|---|---|---|---|---|---|---|
| a) | 14 | 360 | 10 | 18±0.1 | 0.8±0.2 | 57±7 |
| b) | 14 | 360 | 20 | 13±0.1 | 0.63±0.1 | 73±8 |
| c) | 14 | 380 | 20 | 4±0.5 | 1.1±0.25 | 60±10 |
| d) | 58 | 360 | 10 | 3±0.15 | 1.3±0.1 | 68±5 |
| e) | 58 | 360 | 20 | 2±0.1 | 0.94±0.1 | 80±6 |
| f) | 58 | 380 | 20 | < 0.1 | 1±0.25 | 70±8 |

**Table 2.** Summary of the NW growth conditions and the resulting NW density ($d_{NW}$), length ($L_{NW}$) and diameter ($D_{NW}$) deduced from the SEM images in Figure 6.

We now turn to the effect of the InP growth temperature, $T_G$, on the InP NW density. We have fixed the $P_2$ BEP at 1.2x $10^{-5}$ torr, corresponding to a V/III BEP ratio = 20, and varied $T_G$ to 360°C and 380°C. Figure 6 (b,c,e,f) shows the SEM images of the as-grown InP NWs grown with an In/Au BEP ratio = 14 at $T_G$ = 360 °C (b), 380°C (c) and an In/Au BEP ratio = 58 at $T_G$ = 360 °C (e), 380°C (f).

We can observe that when $T_G$ is increased from 360°C to 380°C, the NW density is strongly decreased: from 13 to 4 μm$^{-2}$ and from 2 to < 0.1 μm$^{-2}$, for In/Au BEP ratio = 14 and 58, respectively. Because of the low solubility of phosphorus in the Au droplets, the decrease in $d_{NW}$ when $T_G$ is increased can be attributed to a limited phosphorus supplying to the droplet due to a faster phosphorus desorption at the Au droplet surface. This limited phosphorus supplying can be a factor leading to a limited number of efficient droplets to nucleate InP NWs. Obtaining such a low density (< 0.1 μm$^{-2}$) for InP NWs monolithically grown on Si with these growth conditions ($T_G$ = 380°C and V/III BEP ratio = 20) is of prime importance since they ensure the growth of pure WZ InP NWs [19][20][24]. Here also, it can be observed from Table

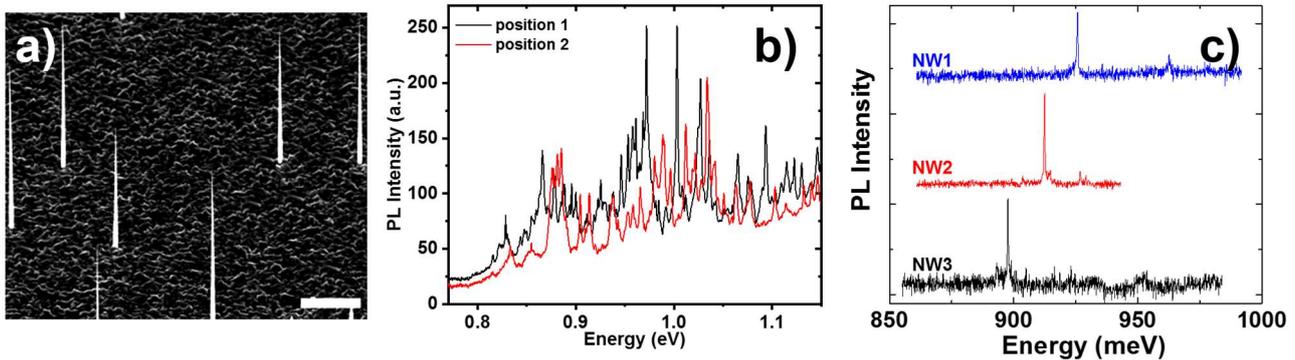

**Figure 7.** a) Typical tilted (45°) SEM image of low-density-needlelike-tapered InAs/InP QD-NWs with $D_{NW}$ = 360 nm at the bottom part of the NW and $L_{NW}$ = 14 μm. Scale bar is 5 μm. b) Low temperature (12 K) PL showing the emission lines from several InAs QDs at two different positions on the sample. c) Low temperature (12 K) micro-PL showing single emission lines from a single InAs/InP QD- NW.

2 a noticeable change in the NW length and diameter when $T_G$ is increased. The increase in the NW length and the decrease in the NW diameter when $T_G$ is increased is explained by a higher diffusion length for the In adatoms on the NW facets towards the Au droplets leading to a higher In supplying of the Au droplets and consequently favoring the axial growth over the radial one.

*3.3 Single InAs/InP QD-NW spectroscopy*

The previous experimental results have shown the possibility to control the density of the InP NWs where ultra-low $d_{NW}$ < 0.1 μm$^{-2}$ was achieved (Figure 6f). To show that this method is effective for the single QD-NW spectroscopy on the as-grown sample, we have used the same growth conditions for the InP NWs as in Figure 6f and grew a single InAs QD inside the NWs. After 12 mins of InP NW growth, the InAs QDs were realized at 420°C for 3 sec with As/In BEP ratio = 20 and then the InP NW growth is continued for 2 mins [24]. Note that the growth time of the InP NWs is the same, 14 mins, compared to that of Figure 6f, resulting in a similar NW diameter, $D_{NW}$ = 70 nm. This $D_{NW}$ is too small to support guided modes and the QD light emission is inhibited by such a NW diameter [30]. In Figure 7a we show the SEM image of low-density-needlelike-tapered InAs/InP QD-NWs with $D_{NW}$ = 360±15 nm at the bottom part of the NWs and $L_{NW}$ ≈ 14 μm. Details on the growth conditions for this QD-NW sample are mentioned elsewhere [24]. With this approach, the light emitted by the InAs QDs will couple to the InP NW HE$_{11}$ waveguide mode thus removing the QD inhibition and enhancing the QD spontaneous emission [31]. Low temperature PL (T = 12 K) in Figure 7b shows the emission spectra of the InAs QDs at two different positions of the sample. Despite the ≈200 μm diameter laser spot, the two spectra show a forest of narrow lines corresponding to single QD emissions as a strong evidence of a low QD-NW density. Moreover, the two spectra exhibit approximately the same number of sharp lines with comparable intensities indicating that the density of the InAs/InP QD-NWs is quite homogenous all over the sample. In addition, micro-PL experiments were performed at 12 K where single QD-NWs are addressed thanks to the low QD-NW density. Single emission lines are observed in the telecom band at low excitation power as shown in Figure 7c. The measured QD linewidths are in the range of 0.4-0.5 meV and are limited by the resolution of the experimental setup. All these experimental results confirm that the achieved QD-NW density is compatible with single QD spectroscopy. As a consequence, we have demonstrated that the NW growth conditions can be optimized to allow single NW spectroscopy directly on the as-grown sample without having to pattern the substrate before the growth or to transfer of NWs on a host substrate.

**4. Conclusion**

We have successfully implemented a new method to control the density of InP NWs monolithically grown on silicon using VLS-MBE without any pre-growth effort. The NW density was tuned between 18 μm$^{-2}$ to values as small as < 0.1 μm$^{-2}$ by investigating different growth parameters. By the variation of the Au BEP used for the Au-In droplet formation, their size and density were changed and such a modification has induced a remarkable change in the NW density. Beyond the Au-In droplet formation, we have extended our study and showed the $P_2$ BEP (or the V/III BEP ratio) could be another driving parameter for the NW density control at the early stage of the growth in addition to the NW growth temperature. Our best result stands out with the use of In/Au BEP ratio = 58, V/III BEP ratio = 20 and $T_G$ = 380°C which are the optimal parameters for ultra-low NW density < 0.1 μm$^{-2}$ while conserving the pure WZ structure of the NWs. This method has been proved effective through the monolithic integration of the InAs/InP QD-NWs on silicon showing a single emission line from a single InAs QD emitting in the telecom band. We believe that the control and understanding reported in this work represents a promise for future integration of III-V single QD-NW devices on silicon.

## Acknowledgements

The authors thank the NanoLyon platform for access to equipments and J. B. Goure and C. Botella for technical assistance.